\documentclass[prl,preprint,lineno,superscriptaddress]{revtex4-2}
\RequirePackage{lineno}
\usepackage[symbol]{footmisc}
\usepackage{graphicx}
\usepackage{dcolumn}
\usepackage{amsmath}
\usepackage{amssymb}
\usepackage{bm}
\usepackage{pifont}
\usepackage{epsfig}
\usepackage{psfrag}
\usepackage{epstopdf}
\usepackage{wasysym}
\usepackage[usenames]{color}
\usepackage{setspace}
\usepackage{hyperref}
\usepackage{wrapfig}
\usepackage[normalem]{ulem}
\usepackage{xcolor}
\usepackage{lineno}

\hypersetup{colorlinks = true, citecolor=blue, linkcolor=blue, urlcolor=blue}

\begin{document}
\title{Intermediate-range order governs dynamics in dense colloidal liquids}
\author{Navneet Singh \footnote{navneet22may@gmail.com}}
\affiliation{Chemistry and Physics of Materials Unit, Jawaharlal Nehru Centre for Advanced Scientific Research, Jakkur, Bangalore - 560064, INDIA}
\author{Zhen Zhang}
\affiliation{State Key Laboratory for Mechanical Behavior of Materials, Xi'an Jiaotong University, Xi'an 710049, China}
\author{A. K. Sood}
\affiliation{Department of Physics, Indian Institute of Science, Bangalore 560012, INDIA}
\affiliation{International Centre for Materials Science, Jawaharlal Nehru Centre for Advanced Scientific Research, Jakkur, Bangalore - 560064, INDIA}
\author{Walter Kob}
\affiliation{Laboratoire Charles Coulomb, University of Montpellier, CNRS, F-34095 Montpellier, France}
\author{Rajesh Ganapathy}
\affiliation{International Centre for Materials Science, Jawaharlal Nehru Centre for Advanced Scientific Research, Jakkur, Bangalore - 560064, INDIA}
\affiliation{School of Advanced Materials (SAMat), Jawaharlal Nehru Centre for Advanced Scientific Research, Jakkur, Bangalore - 560064, INDIA}
\date{\today}

\draft
%\linenumbers
\begin{abstract}
\textbf{The conventional wisdom is that liquids are completely disordered and lack non-trivial structure beyond nearest-neighbor distances. Recent observations have upended this view and demonstrated that the microstructure in liquids is surprisingly rich and plays a critical role in numerous physical, biological, and industrial processes. However, approaches to uncover this structure are either system-specific or yield results that are not physically intuitive. Here, through single-particle resolved three-dimensional confocal microscope imaging and the use of a recently introduced four-point correlation function, we show that bidisperse colloidal liquids have a highly non-trivial structure comprising alternating layers with icosahedral and dodecahedral order, which extends well-beyond nearest-neighbor distances and grows with supercooling. By quantifying the dynamics of the system on the particle level, we establish that it is this intermediate-range order, and not the short-range order, which has a one-to-one correlation with dynamical heterogeneities, a property directly related to the relaxation dynamics of glassy liquids. Our experimental findings provide a direct and much sought-after link between the structure and dynamics of liquids and pave the way for probing the consequences of this intermediate-range order in other liquid state processes.} 
\end{abstract}

\maketitle
\renewcommand{\thefootnote}

\section{Introduction}  
There is growing evidence of liquids having non-trivial structural order extending well-beyond nearest-neighbor separations, and that this ordering plays a decisive role in many fundamental processes, such as liquid-liquid phase transitions \cite{tanaka2022}, dynamical slowing down and the glass transition \cite{frank1952, jonsson1988icosahedral, biroli2008, kob2012non, HimaNagamanasa2015, gokhale2016deconstructing, royall2015role}, as well as nucleation pathway and polymorph selection  during freezing \cite{tanaka2010, kawasaki2010, russo2012, tanaka2022, sun2020crystal}. However, uncovering this subtle structure of liquids requires approaches beyond the standard ensemble-averaged two-point equal time density correlators \cite{hansen2013theory}, which are too coarse a measure. Typically these approaches involve identifying a liquid's locally-favored structure (LFS) - motifs that satisfy a local, but not a global, free energy constraint \cite{frank1952, coslovich2007understanding, royall2015role, Hirata2013, lan2021, Yuan2022}, determining the preferred bond-orientational order (BOO) in liquids \cite{tanaka2010}, or order-agnostic approaches that require computing multipoint spatial correlators \cite{biroli2008, kob2012non, HimaNagamanasa2015, sausset2011characterizing}. Although there is evidence that LFS motifs form larger clusters \cite{royall2015role} and domains of high BOO exist in liquids \cite{kawasaki2010}, and these appear to correlate with dynamics, these approaches are highly system-specific, making it difficult to draw general conclusions about the role of structure in determining the behavior of liquids. In addition, identifying LFS in multi-component and network liquids is usually a daunting task and in practice often impossible. On the other hand, while being independent of system details, the oft-used order-agnostic approaches~\cite{kob2012non} are challenging to implement, if not impossible, in experiments~\cite{HimaNagamanasa2015}, and the structure they identify is difficult to interpret in physical terms.  This last aspect is also one of the drawbacks with machine learning based techniques for probing liquid structure \cite{schoenholz2016, bapst2020}.

A recent advance that sidesteps these issues involves computing a four-point spatial correlation function highly sensitive to angular structural order that may be present in liquids \cite{zhang2020revealing}. Despite being order-agnostic, the strength of this approach lies in its simplicity, unlike the methods mentioned above. While that numerical study revealed that both simple and network liquids do indeed have a very rich angular structural order, extending to distances exceeding eight particle diameters upon supercooling, that study did not address whether the discovered structure correlates with dynamics \cite{zhang2020revealing}.  Although a subsequent experiment on bidisperse granules confirmed the existence of such an intermediate-range order \cite{yuan2021connecting}, the system's athermal and dissipative nature precluded to establish a link between structure and dynamics. Establishing such a link is a long-standing goal in liquid state physics, specifically in the context of the glass transition problem \cite{berthier2011theoretical}.

The only experimental prerequisite to implement the approach laid out in \cite{zhang2020revealing} is access to particle position information in three dimensions (3D). Therefore, colloidal liquids are particularly well-suited systems since single-particle resolution imaging techniques allow to access directly the structure and dynamics in 3D~\cite{gokhale2016deconstructing}. Importantly, since they are also thermalized like real liquids, seeking out a connection between intermediate-range structure, if it does indeed exist, and dynamics becomes possible. Here we address this question by performing fast 3D confocal imaging of bidisperse colloidal liquids. We use a binary mixture of Poly-methylmethacrylate (PMMA) colloidal particles, of radius $r_S = 0.755 $~$\mu$m and $r_L = 0.908 $~$\mu$m $(r_L/r_S \approx 1.20)$, suspended in a mixture of cyclohexyl bromide and decalin, which matches the density and the refractive index of the particles \cite{leunissen2005ionic, leunissen2007manipulating}. The particles were tagged with a fluorophore to enable 3D confocal imaging (see Materials and Methods). We maintained the number density ratio at $ \approx 80:20 $ between large and small particles to prevent crystallization and also to make contact with Ref. \cite{zhang2020revealing}, which used a similar number density ratio \cite{kob1994scaling}. To mitigate wall effects, only particles that were at least $30$ $\mu$m away from the bottom of the sample cell were considered for the analysis \cite{kurita2010experimental, weeks2000three, nagamanasa2011confined}.

\section{Results}	
\subsection*{Evidence for intermediate-range order in colloidal liquids}
\begin{figure*}
\centering
\includegraphics[width=1.0\textwidth]{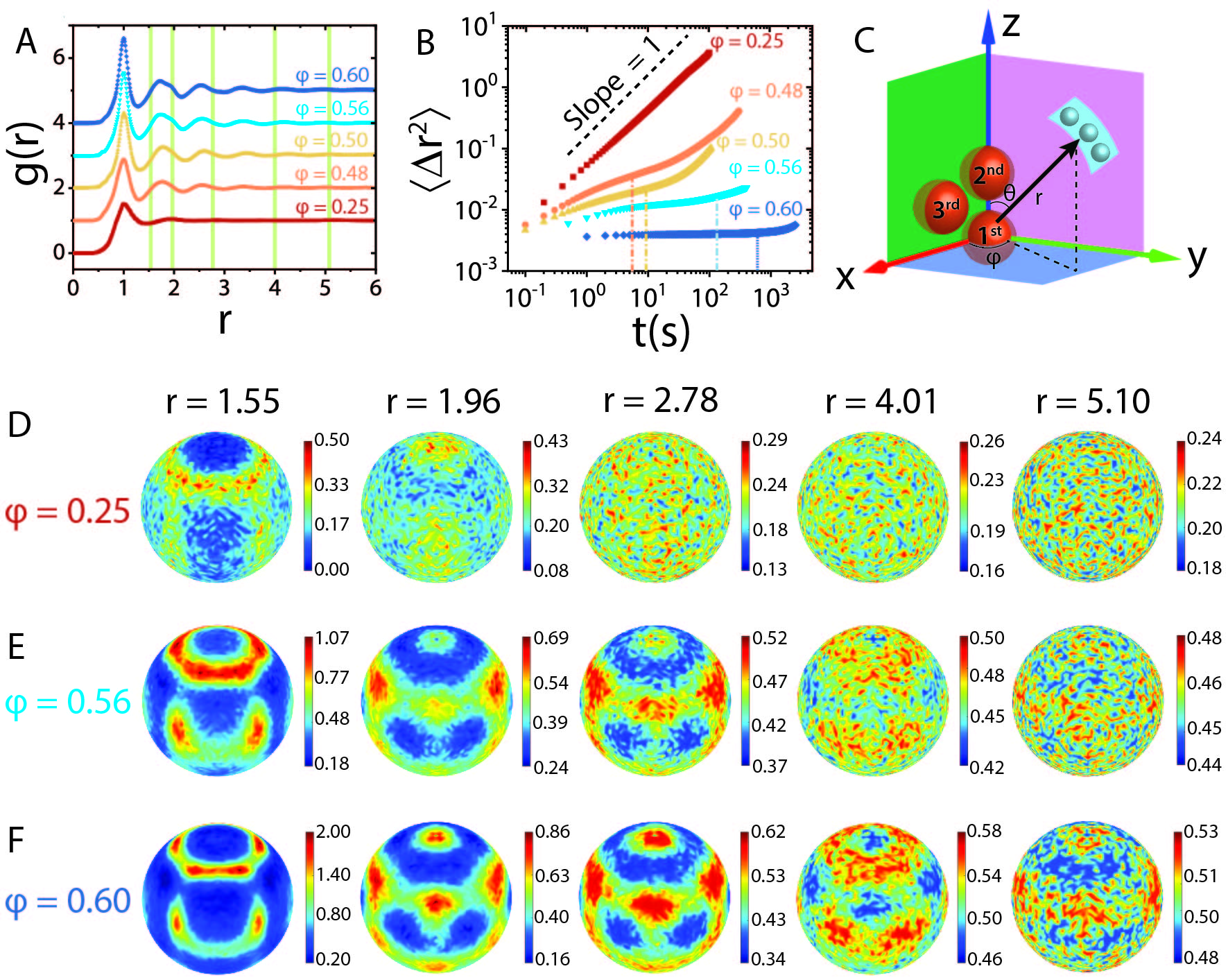}
\caption{\textbf{Three-dimensional distribution of particles in a binary colloidal mixture.} \textbf{(A)} The radial distribution function, $ g(r) $, for various volume fractions $ \varphi $. Different curves have been shifted vertically to improve readability. The radial distance, $ r' $, has been rescaled to $ r = \alpha r'/ \sigma $, where $ \sigma $ is the size of the large particle, and $ \alpha $ is chosen such that the location of the first peak is at $r=1$. The values of $\alpha$ are: 0.70, 0.75, 0.77, 0.80, and 0.84 for $ \varphi = $ 0.25, 0.48, 0.50, 0.56, and 0.60, respectively. \textbf{(B)} The mean squared displacement, $ \langle \Delta r^{2}\rangle $, for various $ \varphi $. The dashed black line has a slope of 1, and the vertical dashed-dotted lines indicate the time scale of maximal non-Gaussianity, $t^{*}$. \textbf{(C)} The local coordinate system consists of three nearest neighbor large particles used to calculate the density distribution, $\rho(\theta, \phi, r)$. \textbf{(D, E, \& F)} Density distribution $ \rho(\theta, \phi, r) $ for different values of $ r $, that is, the distribution of particles around the central particle that are on a sphere of radius $ r $. The volume fractions are 0.25, 0.56, and 0.60 for \textbf{(D)}, \textbf{(E)}, and \textbf{(F)}, respectively. As $ \varphi $ increases, the anisotropy of the density field also increases, i.e.,  $ \rho(\theta, \phi, r) $ depends on the angles $\theta$ and $\phi$ not only at small distances but also at intermediate-length scales.
} 
\label{Fig1}
\end{figure*}

Our experiments focus on the volume fraction range $ 0.25\leq \varphi \leq 0.60 $ in which the system changes with increasing $\varphi$ from a fluid to a dynamically arrested state, see Movies S1 and S2. Figure~\ref{Fig1}A shows the radial distribution function, $ g(r) $, for different $ \varphi $ (Note that here and in the following, we have rescaled the real distance $r'$ to a new distance $r$ such that the first peaks are at the same position \cite{vanderlinden2013}). As is usual for liquids, $ g(r) $ shows no sharp peaks, indicating the absence of long-range crystalline order, nor any discernible growing structural correlation with $ \varphi $ (see Figs.~S1-S3). We characterize the dynamical slowing down by means of the mean squared displacement (MSD), $\langle \Delta r^{2}(t) \rangle $, Fig.~\ref{Fig1}B. As is typical of liquids at $ \varphi = 0.25 $, the MSD shows diffusive dynamics at all times. With increasing packing fraction, $ \varphi \geq 0.48 $, the MSD develops a plateau which becomes increasingly pronounced with $ \varphi $ since particle caging becomes stronger. For the highest density, $ \varphi = 0.60 $, the MSD shows the characteristics of a dynamically frozen state, i.e., an extended plateau.

The inability of $ g(r) $ to capture a significant change of the structure that can rationalize the slowing down of the dynamics is not entirely surprising, given that it only depends on the radial distance $ r $ and hence neglects the angular correlations that may be present in dense liquids. To detect such a correlation, we follow the approach of Ref.~\cite{zhang2020revealing} and select three large nearest-neighbor particles (Fig.~\ref{Fig1}C). One of these particles (`first particle') is chosen to be the origin of the coordinate system, and we define the line joining the first and a second particle as the $z$-axis, and the plane encompassing all three particles as the $z$-$x$ plane. Using this reference frame, we introduce spherical coordinates and calculate the three-dimensional density distribution, $ \rho(\theta, \phi, r) $, by averaging over all combinations of three large nearest-neighbor particles. Thus $ \rho(\theta, \phi, r) $ is a four-point correlation function, since the density at a given point in space is calculated with respect to three particles \cite{zhang2020revealing, yuan2021connecting}, and hence can be expected to give significant insight into the relationship between structure and dynamics.  

Figure~\ref{Fig1}D displays $\rho(\theta,\phi,r)$ at constant $r$
for $ \varphi = 0.25$ and different values of $r$ (see Movie S3). Similar to  $ g(r) $, for $ \varphi = 0.25 $, the angular correlations in $ \rho(\theta, \phi, r) $ do not extend beyond the first nearest neighbor shell, i.e., one has only short-range order. However, with increasing packing fraction, $ \varphi = 0.56 $ (Fig.~\ref{Fig1}E and Movie S4), the angular dependence of $\rho(\theta,\phi,r)$ can be easily detected up to $ r = 4.01 $ and for $ \varphi = 0.60 $ (Fig.~\ref{Fig1}F and  Movie S5), even at $ r = 5.10 $, a distance which corresponds to the sixth nearest-neighbor shell. While established methods of identifying structural order, such as translational and orientational correlation functions, do not signal the presence of a structural order at intermediate length scales \cite{nelson1989polytetrahedral, ernst1991search}, the 3D density distribution $\rho(\theta,\phi,r)$ reveals that glassy colloidal liquids do in fact have such an order. We also note that for $ r = 1.55 $, corresponding to the distance between the first minimum and the second nearest neighbor peak in $ g(r) $, $ \rho(\theta, \phi, r) $ has a highly nontrivial symmetric shape with dodecahedral-like symmetry while for $r=1.96$ $ \rho(\theta, \phi, r) $ has icosahedral symmetry. With increasing $r$ we find an alternating sequence of icosahedral and dodecahedral symmetries, which is related to the fact that  the particles in one shell occupy the depressions formed by the particles in the previous shell, and that an icosahedron is the dual of a dodecahedron and vice versa, see Fig.~\ref{Fig2}D. The radial range in which this alternating symmetry is detected increases with $\varphi$, indicating a growing order with increasing packing fraction. Other order-agnostic methods for quantifying structural correlation in glass-forming liquids, such as patch correlation \cite{Kurchan2010, sausset2011characterizing}, and point-to-set correlation, which are also multi-point correlators \cite{biroli2008, kob2012non}, do not show such clear static correlations extending to intermediate-length scales, demonstrating the usefulness of the present approach.

\subsection*{Quantifying intermediate-range order}
\begin{figure*}
\centering
\includegraphics[width=1.0\textwidth]{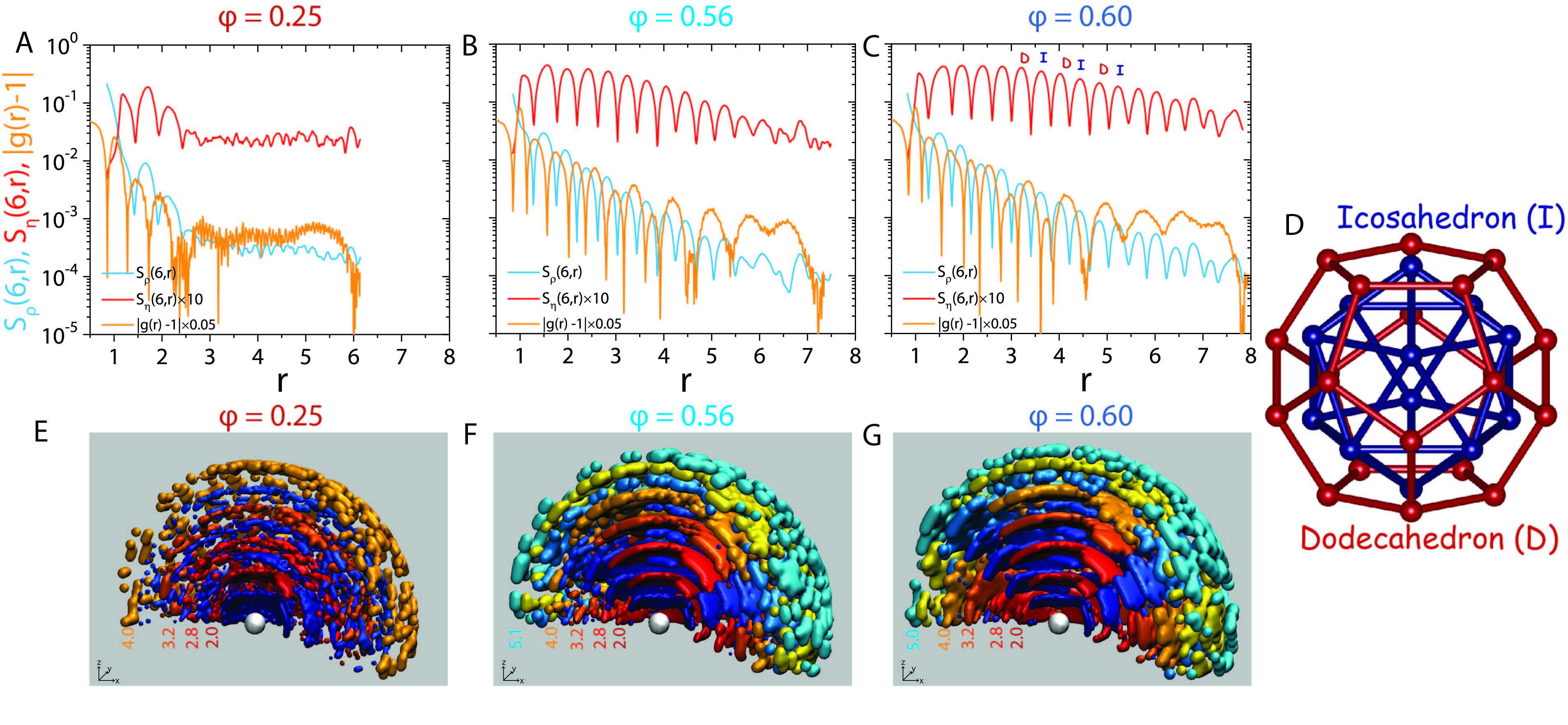}
\caption{\textbf{Quantitative characterization of the density field's structural order and three-dimensional representation.} \textbf{(A)}, \textbf{(B)}, and \textbf{(C)} show angular power spectra, $ S_{\rho}(6, r) $,  normalized angular power spectra, $ S_{\eta}(6, r) $, and radial distribution function, $ |g(r)-1|, $ for $ \varphi = 0.25$, $ \varphi = 0.56$, and $ \varphi = 0.60$, respectively. $ S_{\rho}(6, r) $ (cyan curves) shows an exponential decay as a function of the distance $ r $, whereas $ S_{\eta}(6, r) $ (red curves) stays large even at intermediate $ r $. The local maxima of $ S_{\rho}(6, r) $, and  $ S_{\eta}(6, r) $ are near the local minima of $ |g(r)-1| $ (orange curves). \textbf{(D)} An icosahedron (blue color) is the dual polyhedron of a dodecahedron (red color), with the centers of the faces of one polyhedron corresponding to the vertices of the other. \textbf{(E)}, \textbf{(F)}, and \textbf{(G)} show the three-dimensional representation of the density field for $ \varphi = 0.25$, $ \varphi = 0.56$, and $ \varphi = 0.60$, respectively. The shown layers correspond to distances at which $S_{\rho}(6, r)$ has a local maximum. Only regions with high density (covering $25\%$ surface area of the sphere) are depicted. The bluish/reddish colors correspond to the locations of the alternate local maxima of $S_{\eta}(6, r)$ and thus correspond to shells with icosahedral/dodecahedral symmetry.
} 
	\label{Fig2}
\end{figure*}

Further insight into the observed intermediate-range order is obtained by quantifying the anisotropy of $ \rho(\theta, \phi, r) $. For this, we decompose the angular signal on the sphere into spherical harmonics, $ Y_{m}^{l}(\theta, \phi) $, as $ \rho(\theta,\phi,r) = \sum_{l=0}^{\infty}\sum_{m = -l}^{l} \rho_{l}^{m}(r)Y_{l}^{m}(\theta,\phi) $, and then compute the square root of the angular power spectrum, $ S_{\rho}(l,r) =  \left[ (2l+1)^{-1} \sum_{m = -l}^{l} \left|\rho_{l}^{m}(r)\right|^{2}\right]^{1/2} $, where $ \rho_{l}^{m}(r) $ are the expansion coefficients (see Materials and Methods) \cite{zhang2020revealing}.  To probe these angular correlations in an order agnostic manner, we computed $ S_{\rho}(l,r) $ for various choices of $ l $ (see Figs.~S4-S6). The largest signal is found for $ l = 6 $, in agreement with the icosahedral/dodecahedral symmetries seen in the snapshots of Fig.~\ref{Fig1}. Figures~\ref{Fig2}A-C show the $ r $-dependence of $ S_{\rho}(l = 6, r) $, for $ \varphi = 0.25$, $  0.56 $, and $0.60$, respectively. For all $ \varphi $, the envelope of $S_{\rho}(l = 6, r)$ decays exponentially with $ r $, in agreement with the results of simulations \cite{zhang2020revealing}, with a decay length that increases with $\varphi$. 

Note that $S_{\rho}(l =6,r) $ is sensitive not only to the angular dependence but also to the amplitude of $ \rho(\theta, \phi, r) $. In order to take this into account, we consider the normalized distribution $\eta(\theta,\phi,r) = \frac{\rho(\theta,\phi,r) - \rho_{\text{min}}(\theta,\phi,r)}{\rho_{\text{max}}(\theta,\phi,r) - \rho_{\text{min}}(\theta,\phi,r)}$, where $ \rho_{\text{max}}(\theta,\phi,r) $ and $ \rho_{\text{min}}(\theta,\phi,r) $ represent, respectively, the maximum and minimum of $ \rho(\theta,\phi,r) $ at fixed $ r $ (see Materials and Methods). The square root of the normalized angular power spectrum, $ S_{\eta}(l = 6,r) $, calculated akin to $ S_{\rho}(l = 6, r) $, is included in Fig.~\ref{Fig2}(A)-(C) as well, and reveals that angular correlations are indeed long-range and grow with $\varphi$ (see also Fig.~S7). The simulations of the Lennard-Jones liquid of Ref. \cite{zhang2020revealing} found that the height of the local maxima of $ S_{\eta}(l = 6, r) $ alternates between high and low values. Our $r$-dependence of $S_\eta(6,r)$ does not reveal this and we relate this to the polydispersity of our colloidal particles or the fact that here the interaction potential is much softer than the Lennard-Jones potential considered in Ref.~\cite{zhang2020revealing}. We also note that the height of the first peak of $ S_{\eta}(l = 6, r) $ is smaller than the one of subsequent peaks, suggesting that the icosahedral order (if at all present) is weaker in the first nearest-neighbor shell and is clearly not the locally favoured structure \cite{royall2015role, hallett2018}. 

We also remark that for all $ \varphi $, the decay of $ S_{\rho}(6,r) $ matches very well the one seen in $ g(r) $, which is reasonable since the angular integral of $ \rho (\theta, \phi, r) $ is roughly proportional to $ g(r) $. Interestingly we find that the location of the local maxima of $ S_{\rho}(6,r) $ are near the local minima of $ | g(r) - 1| $ a result that matches the behavior of open network liquids like silica but not the one of the BLJM system  \cite{zhang2020revealing}. This observation demonstrates that $S_\rho(6,r)$ encodes interesting structural information that should be investigated further in the future.

Figure \ref{Fig2}E-G display a 3D representation of the density field for $ \varphi = 0.25$, $0.56 $, and $0.60$, respectively. The layers shown correspond to distances at which $ S_\rho(6,r) $ has a local maximum. The bluish and reddish hues correspond to the alternate local maxima of $ S_{\rho}(6,r) $, which beautifully capture the growing angular anisotropy in liquids with supercooling. These striking visual density maps for $ \varphi = 0.56 $ and $ \varphi = 0.60 $ indicate that liquids are significantly ordered, with the interlocking layers alternatively possessing an icosahedral (bluish color) and dodecahedral symmetry (reddish color).

\subsection*{Connecting intermediate-range order and dynamics}

\begin{figure*}
\centering
\includegraphics[width = 1.0\textwidth]{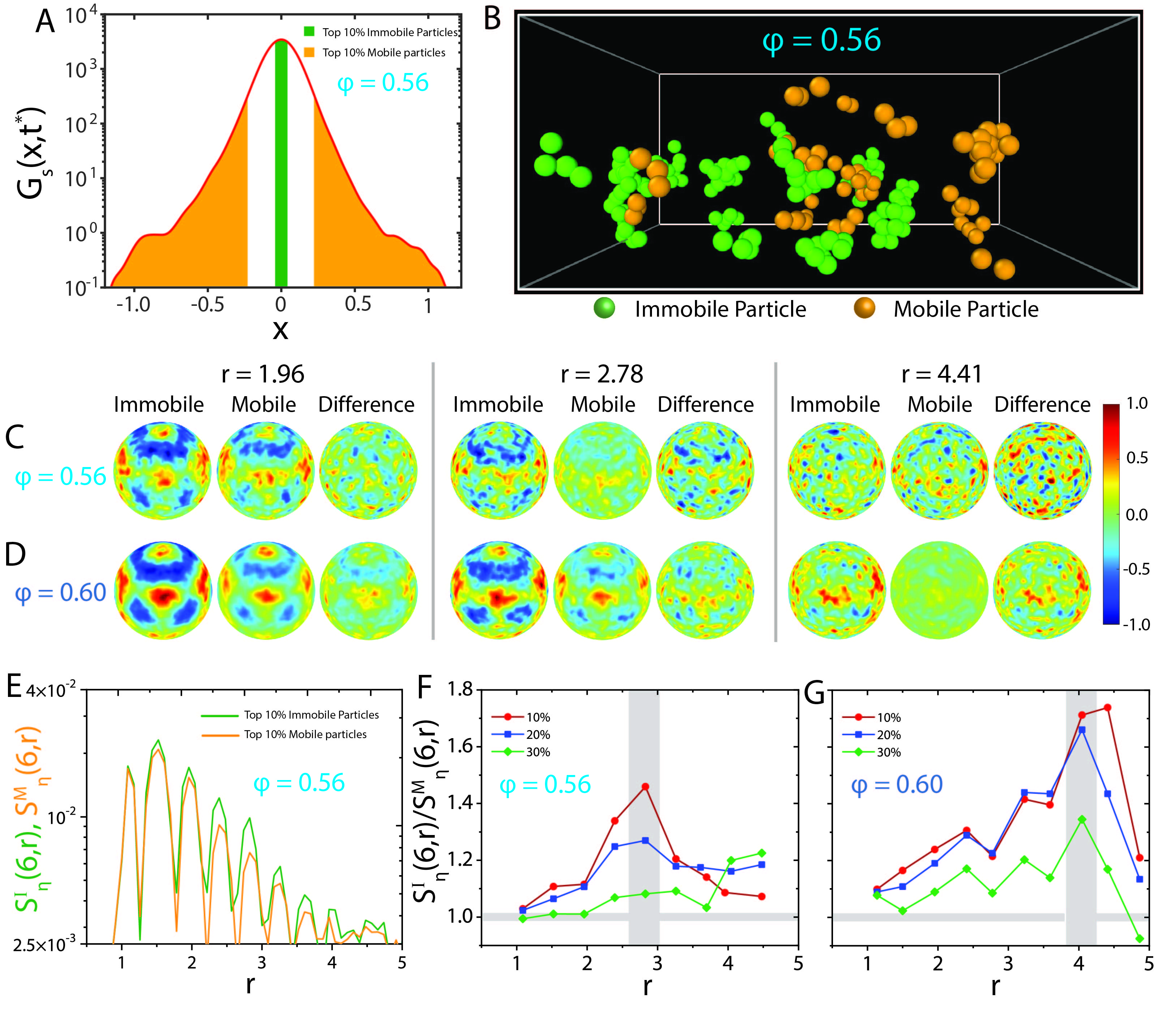}
\caption{\textbf{Probing the connection between intermediate-range order and dynamics.} \textbf{(A)} Probability distribution of particle displacements (Van Hove correlation function), $G_s(x,t)$, at $t = t^{*}$, for $ \varphi = 0.56$. Particles inside the green and orange shaded region are the top 10\% immobile particles and 10\% mobile particles, respectively. \textbf{(B)} The locations of the top 10\% immobile particles (green spheres), and 10\% mobile particles (orange spheres), for $ \varphi = 0.56$. Here we only show mobile and immobile clusters of size $\geq$ 3 particles for ease of viewing. \textbf{(C \& D)} The normalized deviation of the density field from mean density for top 10\% immobile particles, mobile particles, and the difference between the two for different $ r $ values. The plots are normalized using the maximum of the immobile density map for a fixed $ \varphi $ and $ r $. $ \varphi $'s are 0.56 and 0.60 for \textbf{(C)}, and \textbf{(D)}, respectively. The colormaps show immobile particles are more ordered compared to most mobile ones. \textbf{(E)} $ S_{\eta}(6, r) $ for top 10\% immobile ($ S^{I}_{\eta}(6, r) $, green curve) and mobile particles ($ S^{M}_{\eta}(6, r) $, orange curve) versus $ r $ for $ \varphi = 0.56$. \textbf{(F \& G)} shows the non-monotonic evolution of $ S^{I}_{\eta}(6, r)/S^{M}_{\eta}(6, r) $ as a function of $ r $ for $ \varphi = 0.56$, and $ \varphi = 0.60$, respectively. Here mobility threshold cutoffs of 10\%, 20\%, and 30\% are shown in red circles, blue squares, and green diamonds, respectively. The peak position of $ S^{I}_{\eta}(6, r)/S^{M}_{\eta}(6, r) $ shifts from $ r \approx 2.8 $ to $ r \approx 4.0 $ as the volume fraction changes from $ \varphi = 0.56$ to  $ \varphi = 0.60$.}
\label{Fig3}
\end{figure*}

Figures \ref{Fig1} and \ref{Fig2} provide clear evidence for a growing intermediate-range order with increasing $ \varphi $. These results were obtained by averaging over the entire system while glassy liquids are known to have a dynamics which is spatio-temporally heterogeneous \cite{kob1997dynamical, weeks2000three, kegal2000direct, vidalRussell2000, berthier2011theoretical, gokhale2016deconstructing}, and many experimental \cite{watanabe2008direct, hallett2018, ganapathi2021, HimaNagamanasa2015, gokhale2016deconstructing, royall2015role}, and simulation \cite{biroli2008, kob2012non, tanaka2010, berthier2011theoretical, schoenholz2016, bapst2020} studies have found that this dynamics has a strong connection to the underlying structure. In light of these previous results, it is natural to investigate whether the strength of the intermediate-range order is significantly different between mobile and immobile regions, expecting that the structural order is more pronounced in the dynamically immobile regions than in the dynamically mobile ones. To identify these dynamical mobile and immobile regions, we first compute the Van Hove correlation function, $ G_s(x,t) $, at the time $ t^{*} $ at which the non-Gaussian parameter has its maximum, i.e., when the dynamical heterogeneities are most pronounced (see Fig.~S8 and Fig.~\ref{Fig1}B). From the tail and the center of $G_s(x,t^{*})$, we identify the top 10\% of the mobile and immobile particles, respectively, Fig.~\ref{Fig3}A (see Fig. S9 for the 20\% and 30\% threshold). Figure~\ref{Fig3}{B} shows a snapshot of these particles. We then use these mobile and immobile particles to calculate, respectively, the density fields $\rho_M(\theta,\phi,r)$ and $\rho_I(\theta,\phi,r)$ and subsequently the normalized deviation of $ \rho_{\alpha}(\theta, \phi, r) $ from the mean density $ \frac{\rho_\alpha(\theta, \phi, r)  - \overline{ \rho_\alpha(\theta, \phi, r)}}{ \text{max}(\rho_I(\theta, \phi, r))} $, with $\alpha \in \{M,I\}$, and difference $ \rho_{D}(\theta, \phi, r)_{DM} $ (see equation~\ref{eq:9}, Fig.~\ref{Fig3}C-D, Movie S6, and S7 for $ \varphi = 0.56 $, and $ 0.60 $, respectively). (For these calculations, only the particle at the center of the sphere was of mobile/immobile type.) The colormaps in Fig.~\ref{Fig3}C-D show that the structure around immobile particles is clearly more ordered than the one around mobile ones, and it extends to larger distances: For $\varphi = 0.56 $ (Fig.~\ref{Fig3}{C} and Movie S6), the difference in the angular signal can be easily detected up to $ r \approx 3 $ and for $ \varphi = 0.60 $ (Fig.~\ref{Fig3}{D} and Movie S7), even at $ r \approx 4.41 $. This finding demonstrates, thus, for the first time, that in dense colloidal liquids the dynamics of immobile and mobile particles are strongly connected to their intermediate-range order. This result can be quantified by considering $ S_{\eta}(l = 6, r) $ for the mobile ($ S^{M}_{\eta}(6, r) $) and immobile ($ S^{I}_{\eta}(6, r) $) particles, Fig.~\ref{Fig3}E (see Fig. S10 for $ \varphi = 0.60 $). As $ r $ increases, we clearly recognize that ordering is more pronounced in the surrounding of the immobile particles than the one of the mobile particles, especially for intermediate $ r $ values (See Figs.~S10 and S11 for the influence of the threshold, here 10\%). Also remarkable is the fact that at small $r$ there is no noticeable difference between $S_\eta^{M}(6, r) $ and $S_\eta^{I}(6, r) $, which suggests that the mobility of a particle may not be very sensitive to its structure in the nearest-neighbor shell.

To further reinforce these conclusions, we plot in Figs.~\ref{Fig3}{F-G} the ratio of peaks height of $ S^{I}_{\eta}(6, r) $ and $ S^{M}_{\eta}(6, r) $ for two different $ \varphi $ and three different mobility threshold (See Fig.~S12). As expected, when the distinction between the immobile and mobile particles is small (larger mobility thresholds), the ratio $ S^{I}_{\eta}(6, r)/S^{M}_{\eta}(6, r) $ remains close to one, see green diamonds. As the mobility threshold is decreased to 20\% (blue squares) and then to 10\% (red circles) for both the $ \varphi$'s, we observe that $ S^{I}_{\eta}(6, r)/S^{M}_{\eta}(6, r) $ increases significantly and, most remarkably, reaches a maximum for intermediate values of $ r $. This finding clearly demonstrates that intermediate-range order, rather than short-range order, is strongly connected to the dynamics of dense colloidal liquids. As we move further out in $ r $, $ S^{I}_{\eta}(6, r)/S^{M}_{\eta}(6, r) $ drops and becomes $ \approx 1 $, most likely owing to the weakening of the intermediate-range order. This result is in line with a recent study using machine learning \cite{bapst2020}, which found that with increasing supercooling, the dynamics of particles could be predicted accurately only when incorporating structural information that extended up to intermediate-length scales ($ r \approx 5 $). However, this quantity, obtained using graph neural networks, can not be directly translated into physically transparent quantities. In contrast to this approach, results obtained from our simple four-point correlator show that structure and dynamics correlate at intermediate-length scales and provide compelling experimental evidence of growing angular structural correlations with supercooling.

\section{Summary and Conclusions}
Our single-particle resolved experiments reveal that the colloidal liquids studied here have a rich structure comprising alternating icosahedral and dodecahedral layers extending to distances well beyond a few particle diameters. While the standard pair-correlation function for these liquids is practically featureless at these distances, the four-point correlators introduced in Ref.~\cite{zhang2020revealing} easily pick up this intermediate range order. In agreement with numerical simulations, this order becomes more pronounced on increasing the particle number density, i.e., on supercooling. By determining the structure separately in the dynamically fast and slow regions, we show that intermediate-range order is significantly more pronounced in the slow regions than the fast ones, while we find little difference in the structure in the first nearest neighbor shell. This is thus the first direct real space evidence that the structure on intermediate length scales is highly relevant for the relaxation dynamics of glass-formers.

Dynamical heterogeneities are believed to hold the key for understanding the nature of the $\alpha$-relaxation and thus the slowing down of the dynamics with increasing density~\cite{berthier2011theoretical,binder2011glassy}. Our work hints that this key might be encoded in structural features on length scales that are significantly larger than expected so far. Hence in the future, this intermediate range structure should be studied in more detail also for other glass-formers since such investigations will allow to unveil connections between structural features with dynamical quantities like the kinetic fragility of the glass-former. The presented results are also consistent with a physical picture in which a static correlation length increases as one approaches the glass transition \cite{biroli2008, kob2012non, tanaka2010, HimaNagamanasa2015, royall2015role, gokhale2016deconstructing}. Determining if the intermediate-range order observed here also influences other liquid state processes, such as crystal nucleation and growth, is a natural step forward. Here again, colloidal systems will be indispensable \cite{poon2004colloids}. Finally, it is tempting to wonder if the growing static length scale on supercooling found in machine learning approaches \cite{bapst2020} is indeed the one determined by the simple four-point correlation function probed here. Answering this question will allow to advance our understanding of the mechanism leading to the dramatic slowing down of the dynamics of glass-forming systems.

\section{Materials and Methods}
Our system consists of a binary mixture of $ N_S $ small and $ N_L $ large  PMMA (Poly-methylmethacrylate) colloids (density: $ d_{\text{PMMA}} = 1.19$~g/cm$^{3} $; dielectric constant $ \epsilon_{r,\ \text{PMMA}} = 2.6 $; refractive index $ n_{\text{PMMA}} = 1.492 $), of radius $ r_S = 0.755 $ $ \mu $m and $ r_L = 0.908 $ $ \mu $m suspended in an oil mixture of cyclohexyl bromide (CHB; Sigma-Aldrich; density $ d_{\text{CHB}} = 1.336$~g/cm$^{3} $; dielectric constant $ \epsilon_{r,\ \text{CHB}} = 7.92 $; refractive index $ n_{\text{CHB}} = 1.4935 $), and decalin (Spectrochem; density $ d_{\text{decalin}} =  	0.896$~g/cm$^{3} $; dielectric constant $ \epsilon_{r,\ \text{decalin}} = 2.176 $; refractive index $ n_{\text{decalin}} = 1.481 $). The refractive index and density of the oil mixture match those of the particles \cite{leunissen2005ionic, leunissen2007manipulating}. The particle size ratio $ r_L/r_S \approx $ $ 1.2 $ and number density ratio  $ \approx 80:20 $ provide sufficient frustration to prevent crystallization \cite{kob1994scaling, kob1995testing}. Adding 260 $\mu$M Tetrabutylammonium chloride salt (Sigma-Aldrich) significantly screens surface charges on particles. The samples were loaded into a closed cylindrical cell and imaged using a Leica SP8-II confocal microscope (63X oil immersion objective, numerical aperture 1.4, excitation wavelength 488 nm and 552 nm). 3D stacks were recorded every 20~s to 30~s, depending on $ \varphi $. The imaged 3D volume was $ \approx $ $ 92 $ $ \mu $m $ \times $ $ 92 $ $\mu$m $ \times $ $ 45 $ $ \mu $m, which contained $ \approx 50,000$ particles for $ \varphi = 0.60$. For all $ \varphi $, we also carried out 2D imaging in the middle of the 3D sample to capture the dynamics at higher frame rates. The adaptive focus control feature of the microscope was employed to maintain the focus on the observed plane throughout the experiment. Frame rates for 2D imaging varied from $ 10 $ Hz to $ 1 $ Hz, depending on $ \varphi $. The scan field of the 2D area was 92 $\mu$m $\times$ 92 $\mu$m and consisted of $\approx 2100$ particles for $ \varphi = 0.60$. Standard Matlab algorithms \cite{crocker1996methods} were used to generate particle trajectories, and subsequent analysis was performed using in-house developed codes. 

To quantify the intermediate-range order the 3D density distribution $ \rho(\theta, \phi, r) $ on the surface of a sphere was decomposed  into spherical harmonics, $ Y_{l}^{m}(\theta,\phi) $:

\begin{equation}\label{eq:1}
	\rho(\theta,\phi,r) = \sum_{l=0}^{\infty}\sum_{m = -l}^{l} \rho_{l}^{m}(r)Y_{l}^{m}(\theta,\phi) \qquad .
\end{equation}

Here, $ \theta $ is the polar angle, $ \phi $ is the azimuthal angle, and $ \rho_{l}^{m}(r) $ are expansion coefficients of $ \rho(\theta, \phi, r) $ and given by:

\begin{equation}\label{eq:2}
	\rho_{l}^{m}(r) = \int_{0}^{2\pi} d\phi \int_{0}^{\pi}\sin(\theta)\; \rho(\theta,\phi,r)Y_{l}^{m*}(\theta,\phi) d\theta \qquad .
\end{equation} 

\noindent
Here $Y_{l}^{m*}(\theta,\phi)$ is the complex conjugate of the spherical harmonics function of degree $l$ and order $m$. The square root of the angular power spectrum is given by:

\begin{equation}\label{eq:3}
	S_{\rho}(l,r) = \left[\frac{\sum_{m = -l}^{l} \left|\rho_{l}^{m}(r)\right|^{2}}{2l+1}\right]^{1/2}
	\qquad .
\end{equation}

The normalized density distribution, $ \eta(\theta, \phi, r) $, was defined as:

\begin{equation}\label{eq:4}
	\eta(\theta,\phi,r) = \frac{\rho(\theta,\phi,r) - \rho_{\text{min}}(\theta,\phi,r)}{\rho_{\text{max}}(\theta,\phi,r) - \rho_{\text{min}}(\theta,\phi,r)} \qquad .
\end{equation}

\noindent
Here $ \rho_{\text{max}}(\theta,\phi,r) $ and $ \rho_{\text{min}}(\theta,\phi,r) $ are, respectively, the maximum and minimum of $ \rho(\theta,\phi,r) $. Akin to $ \rho(\theta,\phi,r) $, the normalized density distribution, $ \eta(\theta, \phi, r) $, was decomposed  into spherical harmonics, $ Y_{l}^{m}(\theta,\phi) $, i.e.,:

\begin{equation}\label{eq:5}
	\eta(\theta,\phi,r) = \sum_{l=0}^{\infty}\sum_{m = -l}^{l} \rho_{l,\eta}^{m}(r)Y_{l}^{m}(\theta,\phi) \qquad .
\end{equation}

Here $ \rho_{l,\eta}^{m}(r) $ are expansion coefficients of $ \eta(\theta, \phi, r) $ and obtained from:
		
\begin{equation}\label{eq:6}
			\rho_{l,\eta}^{m}(r) = \int_{0}^{2\pi} d\phi \int_{0}^{\pi}\sin(\theta)\; \eta(\theta,\phi,r)Y_{l}^{m*}(\theta,\phi) d\theta \qquad ,
\end{equation} 
		
from which we obtain
		
\begin{equation}\label{eq:7}
	S_{\eta}(l,r) = \left[\frac{\sum_{m = -l}^{l} \left|\rho_{l,\eta}^{m}(r)\right|^{2}}{2l+1}\right]^{1/2} \quad .
\end{equation}		

The normalized deviation of the density distribution for immobile and mobile particles from the mean density, $ \rho_\alpha(\theta, \phi, r)_{DM} $, is calculated as:

\begin{equation}\label{eq:8}
	\rho_\alpha(\theta, \phi, r)_{DM} = \frac{\rho_\alpha(\theta, \phi, r)  - \overline{ \rho_\alpha(\theta, \phi, r)}}{ \text{max}(\rho_I(\theta, \phi, r)}
\end{equation}

%\wkr{Can we simplify this last part a bit by introducing $\rho_\alpha$ with $\alpha \in \{I,M\}$?}

Here $ \rho_\alpha(\theta, \phi, r)$ is the density distribution for immobile particles,  with $\alpha \in \{I,M\}$. The normalized deviation of the density distribution for the difference between immobile and mobile particles from mean density, $ \rho_{D}(\theta, \phi, r)_{DM} $, is calculated as:

\begin{equation}\label{eq:9}
	\rho_{D}(\theta, \phi, r)_{DM} = \frac{\rho_I(\theta, \phi, r) - \rho_M(\theta, \phi, r)  - \overline{ \rho_I(\theta, \phi, r) - \rho_M(\theta, \phi, r)}}{ \text{max}(\rho_I(\theta, \phi, r))}
\end{equation}

\section*{Data Availability}
Raw data are available upon reasonable request from the corresponding author.

\section*{Author Contributions}

N.S., W.K., and R.G. designed research. N.S. and R.G. designed experiments. N.S. performed experiments and analyzed the data. Z.Z. and A.K.S. contributed to project development and provided inputs on the manuscript. N.S., W.K., and R.G. wrote the paper.

\section*{Acknowledgements}
N.S. thanks the Council for Scientific and Industrial Research, INDIA, for a  Senior Research Fellowship. A.K.S. thanks the Department of Science and Technology (DST), Govt. of India, for a Year of Science Fellowship. W.K. is senior member of the Insitut universitaire de France. R.G. thanks DST-Nanomission grant no. SR/NM/TP-25/2016 for financial support.

\section*{Competing Interests}
The authors declare no competing interests.

\bibliography{references}

\begin{thebibliography}{43}%
\makeatletter
\providecommand \@ifxundefined [1]{%
 \@ifx{#1\undefined}
}%
\providecommand \@ifnum [1]{%
 \ifnum #1\expandafter \@firstoftwo
 \else \expandafter \@secondoftwo
 \fi
}%
\providecommand \@ifx [1]{%
 \ifx #1\expandafter \@firstoftwo
 \else \expandafter \@secondoftwo
 \fi
}%
\providecommand \natexlab [1]{#1}%
\providecommand \enquote  [1]{``#1''}%
\providecommand \bibnamefont  [1]{#1}%
\providecommand \bibfnamefont [1]{#1}%
\providecommand \citenamefont [1]{#1}%
\providecommand \href@noop [0]{\@secondoftwo}%
\providecommand \href [0]{\begingroup \@sanitize@url \@href}%
\providecommand \@href[1]{\@@startlink{#1}\@@href}%
\providecommand \@@href[1]{\endgroup#1\@@endlink}%
\providecommand \@sanitize@url [0]{\catcode `\\12\catcode `\$12\catcode
  `\&12\catcode `\#12\catcode `\^12\catcode `\_12\catcode `\%12\relax}%
\providecommand \@@startlink[1]{}%
\providecommand \@@endlink[0]{}%
\providecommand \url  [0]{\begingroup\@sanitize@url \@url }%
\providecommand \@url [1]{\endgroup\@href {#1}{\urlprefix }}%
\providecommand \urlprefix  [0]{URL }%
\providecommand \Eprint [0]{\href }%
\providecommand \doibase [0]{https://doi.org/}%
\providecommand \selectlanguage [0]{\@gobble}%
\providecommand \bibinfo  [0]{\@secondoftwo}%
\providecommand \bibfield  [0]{\@secondoftwo}%
\providecommand \translation [1]{[#1]}%
\providecommand \BibitemOpen [0]{}%
\providecommand \bibitemStop [0]{}%
\providecommand \bibitemNoStop [0]{.\EOS\space}%
\providecommand \EOS [0]{\spacefactor3000\relax}%
\providecommand \BibitemShut  [1]{\csname bibitem#1\endcsname}%
\let\auto@bib@innerbib\@empty
%</preamble>
\bibitem [{\citenamefont {Tanaka}(2020)}]{tanaka2022}%
  \BibitemOpen
  \bibfield  {author} {\bibinfo {author} {\bibfnamefont {H.}~\bibnamefont
  {Tanaka}},\ }\bibfield  {title} {\bibinfo {title} {Liquid–liquid transition
  and polyamorphism},\ }\href {https://doi.org/10.1063/5.0021045} {\bibfield
  {journal} {\bibinfo  {journal} {J. Chem. Phys.}\ }\textbf {\bibinfo {volume}
  {153}},\ \bibinfo {pages} {130901} (\bibinfo {year} {2020})}\BibitemShut
  {NoStop}%
\bibitem [{\citenamefont {Frank}(1952)}]{frank1952}%
  \BibitemOpen
  \bibfield  {author} {\bibinfo {author} {\bibfnamefont {F.~C.}\ \bibnamefont
  {Frank}},\ }\bibfield  {title} {\bibinfo {title} {Supercooling of liquids},\
  }\href {https://doi.org/10.1098/rspa.1952.0194} {\bibfield  {journal}
  {\bibinfo  {journal} {Proc. R. Soc. Lond. A}\ }\textbf {\bibinfo {volume}
  {215}},\ \bibinfo {pages} {43} (\bibinfo {year} {1952})}\BibitemShut
  {NoStop}%
\bibitem [{\citenamefont {J\'onsson}\ and\ \citenamefont
  {Andersen}(1988)}]{jonsson1988icosahedral}%
  \BibitemOpen
  \bibfield  {author} {\bibinfo {author} {\bibfnamefont {H.}~\bibnamefont
  {J\'onsson}}\ and\ \bibinfo {author} {\bibfnamefont {H.~C.}\ \bibnamefont
  {Andersen}},\ }\bibfield  {title} {\bibinfo {title} {Icosahedral ordering in
  the {L}ennard-{J}ones liquid and glass},\ }\href
  {https://doi.org/10.1103/PhysRevLett.60.2295} {\bibfield  {journal} {\bibinfo
   {journal} {Phys. Rev. Lett.}\ }\textbf {\bibinfo {volume} {60}},\ \bibinfo
  {pages} {2295} (\bibinfo {year} {1988})}\BibitemShut {NoStop}%
\bibitem [{\citenamefont {Biroli}\ \emph {et~al.}(2008)\citenamefont {Biroli},
  \citenamefont {Bouchaud}, \citenamefont {Cavagna}, \citenamefont {Grigera},\
  and\ \citenamefont {Verrocchio}}]{biroli2008}%
  \BibitemOpen
  \bibfield  {author} {\bibinfo {author} {\bibfnamefont {G.}~\bibnamefont
  {Biroli}}, \bibinfo {author} {\bibfnamefont {J.-P.}\ \bibnamefont
  {Bouchaud}}, \bibinfo {author} {\bibfnamefont {A.}~\bibnamefont {Cavagna}},
  \bibinfo {author} {\bibfnamefont {T.~S.}\ \bibnamefont {Grigera}},\ and\
  \bibinfo {author} {\bibfnamefont {P.}~\bibnamefont {Verrocchio}},\ }\bibfield
   {title} {\bibinfo {title} {Thermodynamic signature of growing amorphous
  order in glass-forming liquids},\ }\href {https://doi.org/10.1038/nphys1050}
  {\bibfield  {journal} {\bibinfo  {journal} {Nat. Phys.}\ }\textbf {\bibinfo
  {volume} {4}},\ \bibinfo {pages} {771} (\bibinfo {year} {2008})}\BibitemShut
  {NoStop}%
\bibitem [{\citenamefont {Kob}\ \emph {et~al.}(2012)\citenamefont {Kob},
  \citenamefont {Rold{\'a}n-Vargas},\ and\ \citenamefont
  {Berthier}}]{kob2012non}%
  \BibitemOpen
  \bibfield  {author} {\bibinfo {author} {\bibfnamefont {W.}~\bibnamefont
  {Kob}}, \bibinfo {author} {\bibfnamefont {S.}~\bibnamefont
  {Rold{\'a}n-Vargas}},\ and\ \bibinfo {author} {\bibfnamefont
  {L.}~\bibnamefont {Berthier}},\ }\bibfield  {title} {\bibinfo {title}
  {Non-monotonic temperature evolution of dynamic correlations in glass-forming
  liquids},\ }\href {https://doi.org/10.1038/nphys2133} {\bibfield  {journal}
  {\bibinfo  {journal} {Nat. Phys.}\ }\textbf {\bibinfo {volume} {8}},\
  \bibinfo {pages} {164} (\bibinfo {year} {2012})}\BibitemShut {NoStop}%
\bibitem [{\citenamefont {Nagamanasa}\ \emph {et~al.}(2015)\citenamefont
  {Nagamanasa}, \citenamefont {Gokhale}, \citenamefont {K.~Sood},\ and\
  \citenamefont {Ganapathy}}]{HimaNagamanasa2015}%
  \BibitemOpen
  \bibfield  {author} {\bibinfo {author} {\bibfnamefont {H.~K.}\ \bibnamefont
  {Nagamanasa}}, \bibinfo {author} {\bibfnamefont {S.}~\bibnamefont {Gokhale}},
  \bibinfo {author} {\bibfnamefont {A.}~\bibnamefont {K.~Sood}},\ and\ \bibinfo
  {author} {\bibfnamefont {R.}~\bibnamefont {Ganapathy}},\ }\bibfield  {title}
  {\bibinfo {title} {Direct measurements of growing amorphous order and
  non-monotonic dynamic correlations in a colloidal glass-former},\ }\href
  {https://doi.org/10.1038/nphys3289} {\bibfield  {journal} {\bibinfo
  {journal} {Nat. Phys.}\ }\textbf {\bibinfo {volume} {11}},\ \bibinfo {pages}
  {403} (\bibinfo {year} {2015})}\BibitemShut {NoStop}%
\bibitem [{\citenamefont {Gokhale}\ \emph {et~al.}(2016)\citenamefont
  {Gokhale}, \citenamefont {Sood},\ and\ \citenamefont
  {Ganapathy}}]{gokhale2016deconstructing}%
  \BibitemOpen
  \bibfield  {author} {\bibinfo {author} {\bibfnamefont {S.}~\bibnamefont
  {Gokhale}}, \bibinfo {author} {\bibfnamefont {A.~K.}\ \bibnamefont {Sood}},\
  and\ \bibinfo {author} {\bibfnamefont {R.}~\bibnamefont {Ganapathy}},\
  }\bibfield  {title} {\bibinfo {title} {Deconstructing the glass transition
  through critical experiments on colloids},\ }\href
  {https://doi.org/10.1080/00018732.2016.1200832} {\bibfield  {journal}
  {\bibinfo  {journal} {Adv. Phys.}\ }\textbf {\bibinfo {volume} {65}},\
  \bibinfo {pages} {363} (\bibinfo {year} {2016})}\BibitemShut {NoStop}%
\bibitem [{\citenamefont {Royall}\ and\ \citenamefont
  {Williams}(2015)}]{royall2015role}%
  \BibitemOpen
  \bibfield  {author} {\bibinfo {author} {\bibfnamefont {C.~P.}\ \bibnamefont
  {Royall}}\ and\ \bibinfo {author} {\bibfnamefont {S.~R.}\ \bibnamefont
  {Williams}},\ }\bibfield  {title} {\bibinfo {title} {The role of local
  structure in dynamical arrest},\ }\href
  {https://doi.org/https://doi.org/10.1016/j.physrep.2014.11.004} {\bibfield
  {journal} {\bibinfo  {journal} {Phys. Rep.}\ }\textbf {\bibinfo {volume}
  {560}},\ \bibinfo {pages} {1} (\bibinfo {year} {2015})}\BibitemShut {NoStop}%
\bibitem [{\citenamefont {Tanaka}\ \emph {et~al.}(2010)\citenamefont {Tanaka},
  \citenamefont {Kawasaki}, \citenamefont {Shintani},\ and\ \citenamefont
  {Watanabe}}]{tanaka2010}%
  \BibitemOpen
  \bibfield  {author} {\bibinfo {author} {\bibfnamefont {H.}~\bibnamefont
  {Tanaka}}, \bibinfo {author} {\bibfnamefont {T.}~\bibnamefont {Kawasaki}},
  \bibinfo {author} {\bibfnamefont {H.}~\bibnamefont {Shintani}},\ and\
  \bibinfo {author} {\bibfnamefont {K.}~\bibnamefont {Watanabe}},\ }\bibfield
  {title} {\bibinfo {title} {Critical-like behaviour of glass-forming
  liquids},\ }\href {https://doi.org/10.1038/nmat2634} {\bibfield  {journal}
  {\bibinfo  {journal} {Nat. Mater}\ }\textbf {\bibinfo {volume} {9}},\
  \bibinfo {pages} {324} (\bibinfo {year} {2010})}\BibitemShut {NoStop}%
\bibitem [{\citenamefont {Kawasaki}\ and\ \citenamefont
  {Tanaka}(2010)}]{kawasaki2010}%
  \BibitemOpen
  \bibfield  {author} {\bibinfo {author} {\bibfnamefont {T.}~\bibnamefont
  {Kawasaki}}\ and\ \bibinfo {author} {\bibfnamefont {H.}~\bibnamefont
  {Tanaka}},\ }\bibfield  {title} {\bibinfo {title} {Formation of a crystal
  nucleus from liquid},\ }\href {https://doi.org/10.1073/pnas.1001040107}
  {\bibfield  {journal} {\bibinfo  {journal} {Proc. Natl. Acad. Sci. U.S.A.}\
  }\textbf {\bibinfo {volume} {107}},\ \bibinfo {pages} {14036} (\bibinfo
  {year} {2010})}\BibitemShut {NoStop}%
\bibitem [{\citenamefont {Russo}\ and\ \citenamefont
  {Tanaka}(2012)}]{russo2012}%
  \BibitemOpen
  \bibfield  {author} {\bibinfo {author} {\bibfnamefont {J.}~\bibnamefont
  {Russo}}\ and\ \bibinfo {author} {\bibfnamefont {H.}~\bibnamefont {Tanaka}},\
  }\bibfield  {title} {\bibinfo {title} {The microscopic pathway to
  crystallization in supercooled liquids},\ }\href
  {https://doi.org/10.1038/srep00505} {\bibfield  {journal} {\bibinfo
  {journal} {Sci. Rep.}\ }\textbf {\bibinfo {volume} {2}},\ \bibinfo {pages}
  {505} (\bibinfo {year} {2012})}\BibitemShut {NoStop}%
\bibitem [{\citenamefont {Sun}\ and\ \citenamefont
  {Harrowell}(2020)}]{sun2020crystal}%
  \BibitemOpen
  \bibfield  {author} {\bibinfo {author} {\bibfnamefont {G.}~\bibnamefont
  {Sun}}\ and\ \bibinfo {author} {\bibfnamefont {P.}~\bibnamefont
  {Harrowell}},\ }\bibfield  {title} {\bibinfo {title} {Crystal growth rates
  and liquid dynamics at the crossover between stable crystal phases},\ }\href
  {https://doi.org/10.1063/5.0006527} {\bibfield  {journal} {\bibinfo
  {journal} {J. Chem. Phys.}\ }\textbf {\bibinfo {volume} {152}},\ \bibinfo
  {pages} {164505} (\bibinfo {year} {2020})}\BibitemShut {NoStop}%
\bibitem [{\citenamefont {Hansen}\ and\ \citenamefont
  {McDonald}(2013)}]{hansen2013theory}%
  \BibitemOpen
  \bibfield  {author} {\bibinfo {author} {\bibfnamefont {J.~P.}\ \bibnamefont
  {Hansen}}\ and\ \bibinfo {author} {\bibfnamefont {I.~R.}\ \bibnamefont
  {McDonald}},\ }\href
  {https://doi.org/https://doi.org/10.1016/B978-0-12-387032-2.00013-1} {\emph
  {\bibinfo {title} {Theory of Simple Liquids}}},\ \bibinfo {edition} {fourth
  edition}\ ed.\ (\bibinfo  {publisher} {Academic Press},\ \bibinfo {address}
  {Oxford},\ \bibinfo {year} {2013})\BibitemShut {NoStop}%
\bibitem [{\citenamefont {Coslovich}\ and\ \citenamefont
  {Pastore}(2007)}]{coslovich2007understanding}%
  \BibitemOpen
  \bibfield  {author} {\bibinfo {author} {\bibfnamefont {D.}~\bibnamefont
  {Coslovich}}\ and\ \bibinfo {author} {\bibfnamefont {G.}~\bibnamefont
  {Pastore}},\ }\bibfield  {title} {\bibinfo {title} {Understanding fragility
  in supercooled {L}ennard-{J}ones mixtures. {I}. {L}ocally preferred
  structures},\ }\href {https://doi.org/10.1063/1.2773716} {\bibfield
  {journal} {\bibinfo  {journal} {J. Chem. Phys.}\ }\textbf {\bibinfo {volume}
  {127}},\ \bibinfo {pages} {124504} (\bibinfo {year} {2007})}\BibitemShut
  {NoStop}%
\bibitem [{\citenamefont {Hirata}\ \emph {et~al.}(2013)\citenamefont {Hirata},
  \citenamefont {Kang}, \citenamefont {Fujita}, \citenamefont {Klumov},
  \citenamefont {Matsue}, \citenamefont {Kotani}, \citenamefont {Yavari},\ and\
  \citenamefont {Chen}}]{Hirata2013}%
  \BibitemOpen
  \bibfield  {author} {\bibinfo {author} {\bibfnamefont {A.}~\bibnamefont
  {Hirata}}, \bibinfo {author} {\bibfnamefont {L.~J.}\ \bibnamefont {Kang}},
  \bibinfo {author} {\bibfnamefont {T.}~\bibnamefont {Fujita}}, \bibinfo
  {author} {\bibfnamefont {B.}~\bibnamefont {Klumov}}, \bibinfo {author}
  {\bibfnamefont {K.}~\bibnamefont {Matsue}}, \bibinfo {author} {\bibfnamefont
  {M.}~\bibnamefont {Kotani}}, \bibinfo {author} {\bibfnamefont {A.~R.}\
  \bibnamefont {Yavari}},\ and\ \bibinfo {author} {\bibfnamefont {M.~W.}\
  \bibnamefont {Chen}},\ }\bibfield  {title} {\bibinfo {title} {Geometric
  frustration of icosahedron in metallic glasses},\ }\href
  {https://doi.org/10.1126/science.1232450} {\bibfield  {journal} {\bibinfo
  {journal} {Science}\ }\textbf {\bibinfo {volume} {341}},\ \bibinfo {pages}
  {376} (\bibinfo {year} {2013})}\BibitemShut {NoStop}%
\bibitem [{\citenamefont {Lan}\ \emph {et~al.}(2021)\citenamefont {Lan},
  \citenamefont {Zhu}, \citenamefont {Wu}, \citenamefont {Gu}, \citenamefont
  {Zhang}, \citenamefont {Kong}, \citenamefont {Liu}, \citenamefont {Song},
  \citenamefont {Liu}, \citenamefont {Sha}, \citenamefont {Wang}, \citenamefont
  {Liu}, \citenamefont {Liu}, \citenamefont {Wang}, \citenamefont {Liu},
  \citenamefont {Ren},\ and\ \citenamefont {Wang}}]{lan2021}%
  \BibitemOpen
  \bibfield  {author} {\bibinfo {author} {\bibfnamefont {S.}~\bibnamefont
  {Lan}}, \bibinfo {author} {\bibfnamefont {L.}~\bibnamefont {Zhu}}, \bibinfo
  {author} {\bibfnamefont {Z.}~\bibnamefont {Wu}}, \bibinfo {author}
  {\bibfnamefont {L.}~\bibnamefont {Gu}}, \bibinfo {author} {\bibfnamefont
  {Q.}~\bibnamefont {Zhang}}, \bibinfo {author} {\bibfnamefont
  {H.}~\bibnamefont {Kong}}, \bibinfo {author} {\bibfnamefont {J.}~\bibnamefont
  {Liu}}, \bibinfo {author} {\bibfnamefont {R.}~\bibnamefont {Song}}, \bibinfo
  {author} {\bibfnamefont {S.}~\bibnamefont {Liu}}, \bibinfo {author}
  {\bibfnamefont {G.}~\bibnamefont {Sha}}, \bibinfo {author} {\bibfnamefont
  {Y.}~\bibnamefont {Wang}}, \bibinfo {author} {\bibfnamefont {Q.}~\bibnamefont
  {Liu}}, \bibinfo {author} {\bibfnamefont {W.}~\bibnamefont {Liu}}, \bibinfo
  {author} {\bibfnamefont {P.}~\bibnamefont {Wang}}, \bibinfo {author}
  {\bibfnamefont {C.-T.}\ \bibnamefont {Liu}}, \bibinfo {author} {\bibfnamefont
  {Y.}~\bibnamefont {Ren}},\ and\ \bibinfo {author} {\bibfnamefont {X.-L.}\
  \bibnamefont {Wang}},\ }\bibfield  {title} {\bibinfo {title} {A medium-range
  structure motif linking amorphous and crystalline states},\ }\href
  {https://doi.org/10.1038/s41563-021-01011-5} {\bibfield  {journal} {\bibinfo
  {journal} {Nat. Mater}\ }\textbf {\bibinfo {volume} {20}},\ \bibinfo {pages}
  {1347} (\bibinfo {year} {2021})}\BibitemShut {NoStop}%
\bibitem [{\citenamefont {Yuan}\ \emph {et~al.}(2022)\citenamefont {Yuan},
  \citenamefont {Kim}, \citenamefont {Zhou}, \citenamefont {Chang},
  \citenamefont {Zhu}, \citenamefont {Nagaoka}, \citenamefont {Yang},
  \citenamefont {Pham}, \citenamefont {Osher}, \citenamefont {Chen},
  \citenamefont {Ercius}, \citenamefont {Schmid},\ and\ \citenamefont
  {Miao}}]{Yuan2022}%
  \BibitemOpen
  \bibfield  {author} {\bibinfo {author} {\bibfnamefont {Y.}~\bibnamefont
  {Yuan}}, \bibinfo {author} {\bibfnamefont {D.~S.}\ \bibnamefont {Kim}},
  \bibinfo {author} {\bibfnamefont {J.}~\bibnamefont {Zhou}}, \bibinfo {author}
  {\bibfnamefont {D.~J.}\ \bibnamefont {Chang}}, \bibinfo {author}
  {\bibfnamefont {F.}~\bibnamefont {Zhu}}, \bibinfo {author} {\bibfnamefont
  {Y.}~\bibnamefont {Nagaoka}}, \bibinfo {author} {\bibfnamefont
  {Y.}~\bibnamefont {Yang}}, \bibinfo {author} {\bibfnamefont {M.}~\bibnamefont
  {Pham}}, \bibinfo {author} {\bibfnamefont {S.~J.}\ \bibnamefont {Osher}},
  \bibinfo {author} {\bibfnamefont {O.}~\bibnamefont {Chen}}, \bibinfo {author}
  {\bibfnamefont {P.}~\bibnamefont {Ercius}}, \bibinfo {author} {\bibfnamefont
  {A.~K.}\ \bibnamefont {Schmid}},\ and\ \bibinfo {author} {\bibfnamefont
  {J.}~\bibnamefont {Miao}},\ }\bibfield  {title} {\bibinfo {title}
  {Three-dimensional atomic packing in amorphous solids with liquid-like
  structure},\ }\href {https://doi.org/10.1038/s41563-021-01114-z} {\bibfield
  {journal} {\bibinfo  {journal} {Nat. Mater}\ }\textbf {\bibinfo {volume}
  {21}},\ \bibinfo {pages} {95} (\bibinfo {year} {2022})}\BibitemShut {NoStop}%
\bibitem [{\citenamefont {Sausset}\ and\ \citenamefont
  {Levine}(2011)}]{sausset2011characterizing}%
  \BibitemOpen
  \bibfield  {author} {\bibinfo {author} {\bibfnamefont {F.}~\bibnamefont
  {Sausset}}\ and\ \bibinfo {author} {\bibfnamefont {D.}~\bibnamefont
  {Levine}},\ }\bibfield  {title} {\bibinfo {title} {Characterizing order in
  amorphous systems},\ }\href {https://doi.org/10.1103/PhysRevLett.107.045501}
  {\bibfield  {journal} {\bibinfo  {journal} {Phys. Rev. Lett.}\ }\textbf
  {\bibinfo {volume} {107}},\ \bibinfo {pages} {045501} (\bibinfo {year}
  {2011})}\BibitemShut {NoStop}%
\bibitem [{\citenamefont {Schoenholz}\ \emph {et~al.}(2016)\citenamefont
  {Schoenholz}, \citenamefont {Cubuk}, \citenamefont {Sussman}, \citenamefont
  {Kaxiras},\ and\ \citenamefont {Liu}}]{schoenholz2016}%
  \BibitemOpen
  \bibfield  {author} {\bibinfo {author} {\bibfnamefont {S.~S.}\ \bibnamefont
  {Schoenholz}}, \bibinfo {author} {\bibfnamefont {E.~D.}\ \bibnamefont
  {Cubuk}}, \bibinfo {author} {\bibfnamefont {D.~M.}\ \bibnamefont {Sussman}},
  \bibinfo {author} {\bibfnamefont {E.}~\bibnamefont {Kaxiras}},\ and\ \bibinfo
  {author} {\bibfnamefont {A.~J.}\ \bibnamefont {Liu}},\ }\bibfield  {title}
  {\bibinfo {title} {A structural approach to relaxation in glassy liquids},\
  }\href {https://doi.org/10.1038/nphys3644} {\bibfield  {journal} {\bibinfo
  {journal} {Nat. Phys.}\ }\textbf {\bibinfo {volume} {12}},\ \bibinfo {pages}
  {469} (\bibinfo {year} {2016})}\BibitemShut {NoStop}%
\bibitem [{\citenamefont {Bapst}\ \emph {et~al.}(2020)\citenamefont {Bapst},
  \citenamefont {Keck}, \citenamefont {Grabska-Barwi{\'{n}}ska}, \citenamefont
  {Donner}, \citenamefont {Cubuk}, \citenamefont {Schoenholz}, \citenamefont
  {Obika}, \citenamefont {Nelson}, \citenamefont {Back}, \citenamefont
  {Hassabis},\ and\ \citenamefont {Kohli}}]{bapst2020}%
  \BibitemOpen
  \bibfield  {author} {\bibinfo {author} {\bibfnamefont {V.}~\bibnamefont
  {Bapst}}, \bibinfo {author} {\bibfnamefont {T.}~\bibnamefont {Keck}},
  \bibinfo {author} {\bibfnamefont {A.}~\bibnamefont
  {Grabska-Barwi{\'{n}}ska}}, \bibinfo {author} {\bibfnamefont
  {C.}~\bibnamefont {Donner}}, \bibinfo {author} {\bibfnamefont {E.~D.}\
  \bibnamefont {Cubuk}}, \bibinfo {author} {\bibfnamefont {S.~S.}\ \bibnamefont
  {Schoenholz}}, \bibinfo {author} {\bibfnamefont {A.}~\bibnamefont {Obika}},
  \bibinfo {author} {\bibfnamefont {A.~W.~R.}\ \bibnamefont {Nelson}}, \bibinfo
  {author} {\bibfnamefont {T.}~\bibnamefont {Back}}, \bibinfo {author}
  {\bibfnamefont {D.}~\bibnamefont {Hassabis}},\ and\ \bibinfo {author}
  {\bibfnamefont {P.}~\bibnamefont {Kohli}},\ }\bibfield  {title} {\bibinfo
  {title} {Unveiling the predictive power of static structure in glassy
  systems},\ }\href {https://doi.org/10.1038/s41567-020-0842-8} {\bibfield
  {journal} {\bibinfo  {journal} {Nat. Phys.}\ }\textbf {\bibinfo {volume}
  {16}},\ \bibinfo {pages} {448} (\bibinfo {year} {2020})}\BibitemShut
  {NoStop}%
\bibitem [{\citenamefont {Zhang}\ and\ \citenamefont
  {Kob}(2020)}]{zhang2020revealing}%
  \BibitemOpen
  \bibfield  {author} {\bibinfo {author} {\bibfnamefont {Z.}~\bibnamefont
  {Zhang}}\ and\ \bibinfo {author} {\bibfnamefont {W.}~\bibnamefont {Kob}},\
  }\bibfield  {title} {\bibinfo {title} {Revealing the three-dimensional
  structure of liquids using four-point correlation functions},\ }\href
  {https://doi.org/10.1073/pnas.2005638117} {\bibfield  {journal} {\bibinfo
  {journal} {Proc. Natl. Acad. Sci. U.S.A.}\ }\textbf {\bibinfo {volume}
  {117}},\ \bibinfo {pages} {14032} (\bibinfo {year} {2020})}\BibitemShut
  {NoStop}%
\bibitem [{\citenamefont {Yuan}\ \emph {et~al.}(2021)\citenamefont {Yuan},
  \citenamefont {Zhang}, \citenamefont {Kob},\ and\ \citenamefont
  {Wang}}]{yuan2021connecting}%
  \BibitemOpen
  \bibfield  {author} {\bibinfo {author} {\bibfnamefont {H.}~\bibnamefont
  {Yuan}}, \bibinfo {author} {\bibfnamefont {Z.}~\bibnamefont {Zhang}},
  \bibinfo {author} {\bibfnamefont {W.}~\bibnamefont {Kob}},\ and\ \bibinfo
  {author} {\bibfnamefont {Y.}~\bibnamefont {Wang}},\ }\bibfield  {title}
  {\bibinfo {title} {Connecting packing efficiency of binary hard sphere
  systems to their intermediate range structure},\ }\href
  {https://doi.org/10.1103/PhysRevLett.127.278001} {\bibfield  {journal}
  {\bibinfo  {journal} {Phys. Rev. Lett.}\ }\textbf {\bibinfo {volume} {127}},\
  \bibinfo {pages} {278001} (\bibinfo {year} {2021})}\BibitemShut {NoStop}%
\bibitem [{\citenamefont {Berthier}\ and\ \citenamefont
  {Biroli}(2011)}]{berthier2011theoretical}%
  \BibitemOpen
  \bibfield  {author} {\bibinfo {author} {\bibfnamefont {L.}~\bibnamefont
  {Berthier}}\ and\ \bibinfo {author} {\bibfnamefont {G.}~\bibnamefont
  {Biroli}},\ }\bibfield  {title} {\bibinfo {title} {Theoretical perspective on
  the glass transition and amorphous materials},\ }\href
  {https://doi.org/10.1103/RevModPhys.83.587} {\bibfield  {journal} {\bibinfo
  {journal} {Rev. Mod. Phys.}\ }\textbf {\bibinfo {volume} {83}},\ \bibinfo
  {pages} {587} (\bibinfo {year} {2011})}\BibitemShut {NoStop}%
\bibitem [{\citenamefont {Leunissen}\ \emph {et~al.}(2005)\citenamefont
  {Leunissen}, \citenamefont {Christova}, \citenamefont {Hynninen},
  \citenamefont {Royall}, \citenamefont {Campbell}, \citenamefont {Imhof},
  \citenamefont {Dijkstra}, \citenamefont {Van~Roij},\ and\ \citenamefont {van
  Blaaderen}}]{leunissen2005ionic}%
  \BibitemOpen
  \bibfield  {author} {\bibinfo {author} {\bibfnamefont {M.~E.}\ \bibnamefont
  {Leunissen}}, \bibinfo {author} {\bibfnamefont {C.~G.}\ \bibnamefont
  {Christova}}, \bibinfo {author} {\bibfnamefont {A.~P.}\ \bibnamefont
  {Hynninen}}, \bibinfo {author} {\bibfnamefont {C.~P.}\ \bibnamefont
  {Royall}}, \bibinfo {author} {\bibfnamefont {A.~I.}\ \bibnamefont
  {Campbell}}, \bibinfo {author} {\bibfnamefont {A.}~\bibnamefont {Imhof}},
  \bibinfo {author} {\bibfnamefont {M.}~\bibnamefont {Dijkstra}}, \bibinfo
  {author} {\bibfnamefont {R.}~\bibnamefont {Van~Roij}},\ and\ \bibinfo
  {author} {\bibfnamefont {A.}~\bibnamefont {van Blaaderen}},\ }\bibfield
  {title} {\bibinfo {title} {Ionic colloidal crystals of oppositely charged
  particles},\ }\href {https://doi.org/10.1038/nature03946} {\bibfield
  {journal} {\bibinfo  {journal} {Nature}\ }\textbf {\bibinfo {volume} {437}},\
  \bibinfo {pages} {235} (\bibinfo {year} {2005})}\BibitemShut {NoStop}%
\bibitem [{\citenamefont {Leunissen}(2007)}]{leunissen2007manipulating}%
  \BibitemOpen
  \bibfield  {author} {\bibinfo {author} {\bibfnamefont {M.}~\bibnamefont
  {Leunissen}},\ }\emph {\bibinfo {title} {Manipulating colloids with charges
  and electric fields}},\ \href
  {https://dspace.library.uu.nl/handle/1874/19911} {Ph.D. thesis},\ \bibinfo
  {school} {Utrecht University} (\bibinfo {year} {2007})\BibitemShut {NoStop}%
\bibitem [{\citenamefont {Kob}\ and\ \citenamefont
  {Andersen}(1994)}]{kob1994scaling}%
  \BibitemOpen
  \bibfield  {author} {\bibinfo {author} {\bibfnamefont {W.}~\bibnamefont
  {Kob}}\ and\ \bibinfo {author} {\bibfnamefont {H.~C.}\ \bibnamefont
  {Andersen}},\ }\bibfield  {title} {\bibinfo {title} {Scaling behavior in the
  $\ensuremath{\beta}$-relaxation regime of a supercooled {L}ennard-{J}ones
  mixture},\ }\href {https://doi.org/10.1103/PhysRevLett.73.1376} {\bibfield
  {journal} {\bibinfo  {journal} {Phys. Rev. Lett.}\ }\textbf {\bibinfo
  {volume} {73}},\ \bibinfo {pages} {1376} (\bibinfo {year}
  {1994})}\BibitemShut {NoStop}%
\bibitem [{\citenamefont {Kurita}\ and\ \citenamefont
  {Weeks}(2010)}]{kurita2010experimental}%
  \BibitemOpen
  \bibfield  {author} {\bibinfo {author} {\bibfnamefont {R.}~\bibnamefont
  {Kurita}}\ and\ \bibinfo {author} {\bibfnamefont {E.~R.}\ \bibnamefont
  {Weeks}},\ }\bibfield  {title} {\bibinfo {title} {Experimental study of
  random-close-packed colloidal particles},\ }\href
  {https://doi.org/10.1103/PhysRevE.82.011403} {\bibfield  {journal} {\bibinfo
  {journal} {Phys. Rev. E}\ }\textbf {\bibinfo {volume} {82}},\ \bibinfo
  {pages} {011403} (\bibinfo {year} {2010})}\BibitemShut {NoStop}%
\bibitem [{\citenamefont {Weeks}\ \emph {et~al.}(2000)\citenamefont {Weeks},
  \citenamefont {Crocker}, \citenamefont {Levitt}, \citenamefont {Schofield},\
  and\ \citenamefont {Weitz}}]{weeks2000three}%
  \BibitemOpen
  \bibfield  {author} {\bibinfo {author} {\bibfnamefont {E.~R.}\ \bibnamefont
  {Weeks}}, \bibinfo {author} {\bibfnamefont {J.~C.}\ \bibnamefont {Crocker}},
  \bibinfo {author} {\bibfnamefont {A.~C.}\ \bibnamefont {Levitt}}, \bibinfo
  {author} {\bibfnamefont {A.}~\bibnamefont {Schofield}},\ and\ \bibinfo
  {author} {\bibfnamefont {D.~A.}\ \bibnamefont {Weitz}},\ }\bibfield  {title}
  {\bibinfo {title} {Three-dimensional direct imaging of structural relaxation
  near the colloidal glass transition},\ }\href
  {https://doi.org/10.1126/science.287.5453.627} {\bibfield  {journal}
  {\bibinfo  {journal} {Science}\ }\textbf {\bibinfo {volume} {287}},\ \bibinfo
  {pages} {627} (\bibinfo {year} {2000})}\BibitemShut {NoStop}%
\bibitem [{\citenamefont {Nagamanasa}\ \emph {et~al.}(2011)\citenamefont
  {Nagamanasa}, \citenamefont {Gokhale}, \citenamefont {Ganapathy},\ and\
  \citenamefont {Sood}}]{nagamanasa2011confined}%
  \BibitemOpen
  \bibfield  {author} {\bibinfo {author} {\bibfnamefont {H.~K.}\ \bibnamefont
  {Nagamanasa}}, \bibinfo {author} {\bibfnamefont {S.}~\bibnamefont {Gokhale}},
  \bibinfo {author} {\bibfnamefont {R.}~\bibnamefont {Ganapathy}},\ and\
  \bibinfo {author} {\bibfnamefont {A.~K.}\ \bibnamefont {Sood}},\ }\bibfield
  {title} {\bibinfo {title} {Confined glassy dynamics at grain boundaries in
  colloidal crystals},\ }\href {https://doi.org/10.1073/pnas.1101858108}
  {\bibfield  {journal} {\bibinfo  {journal} {Proc. Natl. Acad. Sci. U.S.A.}\
  }\textbf {\bibinfo {volume} {108}},\ \bibinfo {pages} {11323} (\bibinfo
  {year} {2011})}\BibitemShut {NoStop}%
\bibitem [{\citenamefont {van~der Linden}\ \emph {et~al.}(2013)\citenamefont
  {van~der Linden}, \citenamefont {El~Masri}, \citenamefont {Dijkstra},\ and\
  \citenamefont {van Blaaderen}}]{vanderlinden2013}%
  \BibitemOpen
  \bibfield  {author} {\bibinfo {author} {\bibfnamefont {M.~N.}\ \bibnamefont
  {van~der Linden}}, \bibinfo {author} {\bibfnamefont {D.}~\bibnamefont
  {El~Masri}}, \bibinfo {author} {\bibfnamefont {M.}~\bibnamefont {Dijkstra}},\
  and\ \bibinfo {author} {\bibfnamefont {A.}~\bibnamefont {van Blaaderen}},\
  }\bibfield  {title} {\bibinfo {title} {Expansion of charged colloids after
  centrifugation: formation and crystallisation of long-range repulsive
  glasses},\ }\href {https://doi.org/10.1039/C3SM51752G} {\bibfield  {journal}
  {\bibinfo  {journal} {Soft Matter}\ }\textbf {\bibinfo {volume} {9}},\
  \bibinfo {pages} {11618} (\bibinfo {year} {2013})}\BibitemShut {NoStop}%
\bibitem [{\citenamefont {Nelson}\ and\ \citenamefont
  {Spaepen}(1989)}]{nelson1989polytetrahedral}%
  \BibitemOpen
  \bibfield  {author} {\bibinfo {author} {\bibfnamefont {D.~R.}\ \bibnamefont
  {Nelson}}\ and\ \bibinfo {author} {\bibfnamefont {F.}~\bibnamefont
  {Spaepen}},\ }\bibfield  {title} {\bibinfo {title} {Polytetrahedral order in
  condensed matter},\ }\bibfield  {booktitle} {\emph {\bibinfo {booktitle}
  {Superconductivity Quasicrystals two Dimensional Physics}},\ }\href
  {https://doi.org/https://doi.org/10.1016/S0081-1947(08)60079-X} {\ \bibinfo
  {series} {Solid State Physics},\ \textbf {\bibinfo {volume} {42}},\ \bibinfo
  {pages} {1} (\bibinfo {year} {1989})}\BibitemShut {NoStop}%
\bibitem [{\citenamefont {Ernst}\ \emph {et~al.}(1991)\citenamefont {Ernst},
  \citenamefont {Nagel},\ and\ \citenamefont {Grest}}]{ernst1991search}%
  \BibitemOpen
  \bibfield  {author} {\bibinfo {author} {\bibfnamefont {R.~M.}\ \bibnamefont
  {Ernst}}, \bibinfo {author} {\bibfnamefont {S.~R.}\ \bibnamefont {Nagel}},\
  and\ \bibinfo {author} {\bibfnamefont {G.~S.}\ \bibnamefont {Grest}},\
  }\bibfield  {title} {\bibinfo {title} {Search for a correlation length in a
  simulation of the glass transition},\ }\href
  {https://doi.org/10.1103/PhysRevB.43.8070} {\bibfield  {journal} {\bibinfo
  {journal} {Phys. Rev. B}\ }\textbf {\bibinfo {volume} {43}},\ \bibinfo
  {pages} {8070} (\bibinfo {year} {1991})}\BibitemShut {NoStop}%
\bibitem [{\citenamefont {Kurchan}\ and\ \citenamefont
  {Levine}(2010)}]{Kurchan2010}%
  \BibitemOpen
  \bibfield  {author} {\bibinfo {author} {\bibfnamefont {J.}~\bibnamefont
  {Kurchan}}\ and\ \bibinfo {author} {\bibfnamefont {D.}~\bibnamefont
  {Levine}},\ }\bibfield  {title} {\bibinfo {title} {Order in glassy systems},\
  }\href {https://doi.org/10.1088/1751-8113/44/3/035001} {\bibfield  {journal}
  {\bibinfo  {journal} {J. Phys. A Math. Theor.}\ }\textbf {\bibinfo {volume}
  {44}},\ \bibinfo {pages} {035001} (\bibinfo {year} {2010})}\BibitemShut
  {NoStop}%
\bibitem [{\citenamefont {Hallett}\ \emph {et~al.}(2018)\citenamefont
  {Hallett}, \citenamefont {Turci},\ and\ \citenamefont
  {Royall}}]{hallett2018}%
  \BibitemOpen
  \bibfield  {author} {\bibinfo {author} {\bibfnamefont {J.~E.}\ \bibnamefont
  {Hallett}}, \bibinfo {author} {\bibfnamefont {F.}~\bibnamefont {Turci}},\
  and\ \bibinfo {author} {\bibfnamefont {C.~P.}\ \bibnamefont {Royall}},\
  }\bibfield  {title} {\bibinfo {title} {Local structure in deeply supercooled
  liquids exhibits growing lengthscales and dynamical correlations},\ }\href
  {https://doi.org/10.1038/s41467-018-05371-6} {\bibfield  {journal} {\bibinfo
  {journal} {Nat. Commun.}\ }\textbf {\bibinfo {volume} {9}},\ \bibinfo {pages}
  {3272} (\bibinfo {year} {2018})}\BibitemShut {NoStop}%
\bibitem [{\citenamefont {Kob}\ \emph {et~al.}(1997)\citenamefont {Kob},
  \citenamefont {Donati}, \citenamefont {Plimpton}, \citenamefont {Poole},\
  and\ \citenamefont {Glotzer}}]{kob1997dynamical}%
  \BibitemOpen
  \bibfield  {author} {\bibinfo {author} {\bibfnamefont {W.}~\bibnamefont
  {Kob}}, \bibinfo {author} {\bibfnamefont {C.}~\bibnamefont {Donati}},
  \bibinfo {author} {\bibfnamefont {S.~J.}\ \bibnamefont {Plimpton}}, \bibinfo
  {author} {\bibfnamefont {P.~H.}\ \bibnamefont {Poole}},\ and\ \bibinfo
  {author} {\bibfnamefont {S.~C.}\ \bibnamefont {Glotzer}},\ }\bibfield
  {title} {\bibinfo {title} {Dynamical heterogeneities in a supercooled
  {L}ennard-{J}ones liquid},\ }\href
  {https://doi.org/10.1103/PhysRevLett.79.2827} {\bibfield  {journal} {\bibinfo
   {journal} {Phys. Rev. Lett.}\ }\textbf {\bibinfo {volume} {79}},\ \bibinfo
  {pages} {2827} (\bibinfo {year} {1997})}\BibitemShut {NoStop}%
\bibitem [{\citenamefont {Kegel}\ and\ \citenamefont {van
  Blaaderen}(2000)}]{kegal2000direct}%
  \BibitemOpen
  \bibfield  {author} {\bibinfo {author} {\bibfnamefont {W.~K.}\ \bibnamefont
  {Kegel}}\ and\ \bibinfo {author} {\bibfnamefont {A.}~\bibnamefont {van
  Blaaderen}},\ }\bibfield  {title} {\bibinfo {title} {Direct observation of
  dynamical heterogeneities in colloidal hard-sphere suspensions},\ }\href
  {https://doi.org/10.1126/science.287.5451.290} {\bibfield  {journal}
  {\bibinfo  {journal} {Science}\ }\textbf {\bibinfo {volume} {287}},\ \bibinfo
  {pages} {290} (\bibinfo {year} {2000})}\BibitemShut {NoStop}%
\bibitem [{\citenamefont {Russell}\ and\ \citenamefont
  {Israeloff}(2000)}]{vidalRussell2000}%
  \BibitemOpen
  \bibfield  {author} {\bibinfo {author} {\bibfnamefont {E.~V.}\ \bibnamefont
  {Russell}}\ and\ \bibinfo {author} {\bibfnamefont {N.~E.}\ \bibnamefont
  {Israeloff}},\ }\bibfield  {title} {\bibinfo {title} {Direct observation of
  molecular cooperativity near the glass transition},\ }\href
  {https://doi.org/10.1038/35047037} {\bibfield  {journal} {\bibinfo  {journal}
  {Nature}\ }\textbf {\bibinfo {volume} {408}},\ \bibinfo {pages} {695}
  (\bibinfo {year} {2000})}\BibitemShut {NoStop}%
\bibitem [{\citenamefont {Watanabe}\ and\ \citenamefont
  {Tanaka}(2008)}]{watanabe2008direct}%
  \BibitemOpen
  \bibfield  {author} {\bibinfo {author} {\bibfnamefont {K.}~\bibnamefont
  {Watanabe}}\ and\ \bibinfo {author} {\bibfnamefont {H.}~\bibnamefont
  {Tanaka}},\ }\bibfield  {title} {\bibinfo {title} {Direct observation of
  medium-range crystalline order in granular liquids near the glass
  transition},\ }\href {https://doi.org/10.1103/PhysRevLett.100.158002}
  {\bibfield  {journal} {\bibinfo  {journal} {Phys. Rev. Lett.}\ }\textbf
  {\bibinfo {volume} {100}},\ \bibinfo {pages} {158002} (\bibinfo {year}
  {2008})}\BibitemShut {NoStop}%
\bibitem [{\citenamefont {Ganapathi}\ \emph {et~al.}(2021)\citenamefont
  {Ganapathi}, \citenamefont {Chakrabarti}, \citenamefont {Sood},\ and\
  \citenamefont {Ganapathy}}]{ganapathi2021}%
  \BibitemOpen
  \bibfield  {author} {\bibinfo {author} {\bibfnamefont {D.}~\bibnamefont
  {Ganapathi}}, \bibinfo {author} {\bibfnamefont {D.}~\bibnamefont
  {Chakrabarti}}, \bibinfo {author} {\bibfnamefont {A.~K.}\ \bibnamefont
  {Sood}},\ and\ \bibinfo {author} {\bibfnamefont {R.}~\bibnamefont
  {Ganapathy}},\ }\bibfield  {title} {\bibinfo {title} {Structure determines
  where crystallization occurs in a soft colloidal glass},\ }\href
  {https://doi.org/10.1038/s41567-020-1016-4} {\bibfield  {journal} {\bibinfo
  {journal} {Nat. Phys.}\ }\textbf {\bibinfo {volume} {17}},\ \bibinfo {pages}
  {114} (\bibinfo {year} {2021})}\BibitemShut {NoStop}%
\bibitem [{\citenamefont {Binder}\ and\ \citenamefont
  {Kob}(2011)}]{binder2011glassy}%
  \BibitemOpen
  \bibfield  {author} {\bibinfo {author} {\bibfnamefont {K.}~\bibnamefont
  {Binder}}\ and\ \bibinfo {author} {\bibfnamefont {W.}~\bibnamefont {Kob}},\
  }\href {https://doi.org/10.1142/7300} {\emph {\bibinfo {title} {Glassy
  materials and disordered solids: An introduction to their statistical
  mechanics}}}\ (\bibinfo  {publisher} {World {S}cientific},\ \bibinfo {year}
  {2011})\BibitemShut {NoStop}%
\bibitem [{\citenamefont {Poon}(2004)}]{poon2004colloids}%
  \BibitemOpen
  \bibfield  {author} {\bibinfo {author} {\bibfnamefont {W.}~\bibnamefont
  {Poon}},\ }\bibfield  {title} {\bibinfo {title} {Colloids as big atoms},\
  }\href {https://doi.org/10.1126/science.1097964} {\bibfield  {journal}
  {\bibinfo  {journal} {Science}\ }\textbf {\bibinfo {volume} {304}},\ \bibinfo
  {pages} {830} (\bibinfo {year} {2004})}\BibitemShut {NoStop}%
\bibitem [{\citenamefont {Kob}\ and\ \citenamefont
  {Andersen}(1995)}]{kob1995testing}%
  \BibitemOpen
  \bibfield  {author} {\bibinfo {author} {\bibfnamefont {W.}~\bibnamefont
  {Kob}}\ and\ \bibinfo {author} {\bibfnamefont {H.~C.}\ \bibnamefont
  {Andersen}},\ }\bibfield  {title} {\bibinfo {title} {Testing mode-coupling
  theory for a supercooled binary {L}ennard-{J}ones mixture {I}: The van {H}ove
  correlation function},\ }\href {https://doi.org/10.1103/PhysRevE.51.4626}
  {\bibfield  {journal} {\bibinfo  {journal} {Phys. Rev. E}\ }\textbf {\bibinfo
  {volume} {51}},\ \bibinfo {pages} {4626} (\bibinfo {year}
  {1995})}\BibitemShut {NoStop}%
\bibitem [{\citenamefont {Crocker}\ and\ \citenamefont
  {Grier}(1996)}]{crocker1996methods}%
  \BibitemOpen
  \bibfield  {author} {\bibinfo {author} {\bibfnamefont {J.~C.}\ \bibnamefont
  {Crocker}}\ and\ \bibinfo {author} {\bibfnamefont {D.~G.}\ \bibnamefont
  {Grier}},\ }\bibfield  {title} {\bibinfo {title} {Methods of digital video
  microscopy for colloidal studies},\ }\href
  {https://doi.org/https://doi.org/10.1006/jcis.1996.0217} {\bibfield
  {journal} {\bibinfo  {journal} {J. Colloid Interface Sci.}\ }\textbf
  {\bibinfo {volume} {179}},\ \bibinfo {pages} {298} (\bibinfo {year}
  {1996})}\BibitemShut {NoStop}%
\end{thebibliography}%

\newpage

%\newpage

%\newpage

\end{document}